\pdfoutput=1

\RequirePackage{fix-cm}
\documentclass[smallextended]{svjour3}       
\smartqed  
\usepackage{graphics,epstopdf}
\usepackage{graphicx}
\usepackage[colorlinks=true,citecolor=blue, urlcolor=blue]{hyperref}
\usepackage{float}

\begin{document}
\title{Local deterministic simulation of equatorial Von Neumann measurements on tripartite GHZ state}
\author{Arup Roy \and
        Amit Mukherjee \and 
        Some Sankar Bhattacharya \and
        Manik Banik \and
        Subhadipa Das
}
\institute{Arup Roy \at
              Physics and Applied Mathematics Unit, Indian Statistical Institute, 203 B.T. Road, Kolkata-700108, India\\
              \email{arup145.roy@gmail.com}
              \and
               Amit Mukherjee \at
                            Physics and Applied Mathematics Unit, Indian Statistical Institute, 203 B.T. Road, Kolkata-700108, India\\
                            \email{amitisiphys@gmail.com}
                           \and
             Some Sankar Bhattacharya\at
                           Physics and Applied Mathematics Unit, Indian Statistical Institute, 203 B.T. Road, Kolkata-700108, India\\
                           \email{somesankar@gmail.com}
                           \and               
            Manik Banik \at
              Physics and Applied Mathematics Unit, Indian Statistical Institute, 203 B.T. Road, Kolkata-700108, India\\
              \email{manik11ju@gmail.com}           
           \and
          Subhadipa Das\at
              S.N. Bose National Center for Basic Sciences, Block JD, Sector III, Salt Lake, Kolkata-700098, India.\\
                            \email{sbhdpa.das@bose.res.in}           
}

\date{Received: date / Accepted: date}
\maketitle

\begin{abstract}
Experimental free-will or measurement independence is one of the crucial assumptions in derivation of any nonlocal theorem. Any nonlocal correlation obtained in quantum world can have a local deterministic explanation if there is no experimental free-will in choosing the measurement settings. Recently, in [\href{http://link.aps.org/doi/10.1103/PhysRevLett.105.250404}{Phys. Rev. Lett. {\bf105}, 250404 (2010)}] it has been shown that to obtain a local deterministic description for singlet state correlation one does not need to give up measurement independence completely, but a partial measurement dependence suffices. In three party scenario considering GHZ correlation one can exhibit absolute contradiction between quantum theory and local realism. In this paper we show that such correlation also has local deterministic description if measurement independence is given up, even if not completely. We provide a local deterministic model for equatorial Von Neumann measurements on tripartite GHZ state by sacrificing measurement independence partially.      
\end{abstract}

\maketitle
\section{Introduction}
One of the most surprising features of quantum mechanics is that it exhibits nonlocal correlations i.e. measurements performed on several quantum systems in an entangled state may contain correlations in their outcomes that cannot be simulated by shared local variables. Though the observation that quantum theory predicts nonlocal correlations goes all the way back to the famous EPR argument \cite{EPR}, it is the remarkable Bell's theorem in 1964 \cite{Bell} which established that the conflict between quantum theory and \emph{local realism} could be experimentally decided and this theorem motivated successful experimental tests in this regard \cite{Aspect1,Aspect2,Aspect3}. Whereas tests of Bell's inequality only demonstrate the said contradictions in a statistical manner, it was the scientist-trio Greenberger, Horne and Zeilinger who first pointed out that considering more than two particles it is possible to demonstrate absolute contradiction between the predictions of local realism and those of quantum mechanics \cite{GHZ} which is experimentally verified in \cite{Pan}. Other than this foundational interest the study of nonlocal correlation has been boosted in the last decade from information theoretic perspective (see \cite{Brunner} for review on Bell nonlocality). It has been proved that nonlocal correlations play important role in ``device independent'' quantum key distribution \cite{Barrett1,Acin,Pironio1,Vazirani} and random number generation \cite{Colbeck,Pironio2,Manik}. Thus quantification of the amount of nonlocality in a correlation demands interest from both the foundational as well as practical perspectives. 

One practical measure to quantify the amount of nonlocality has been introduced in the Refs. \cite{Maudlin,Brassard,Steiner}. They pointed out that the amount of communication that is required (either in the worst case scenario or on average), in addition to shared randomness (a local resource), in order to simulate the behavior of entangled quantum systems could be a good measure of nonlocality. For local correlations, no communication is needed, as shared randomness suffice; they thus have a ``communication measure'' of nonlocality equal to zero. For the bipartite scenario one of the most important result in this regard has been proved by Toner and Bacon, who showed that for the case of local projective measurements on an entangled Bell state exact simulation is possible using local hidden variables augmented by just one bit of classical communication \cite{Toner}. In multi-party scenario investigation of such simulation protocol for Greenberger-Horne-Zeilinger (GHZ) state has been started in \cite{Caves,Chouba}. Branciard and  Gisin proved that 3 classical bits (in total) turn out to be sufficient to simulate all equatorial Von Neumann measurements on the tripartite GHZ state \cite{Branciard}. Recently, Brassard et al \cite{Gilles} have shown that the GHZ joint discrete probability distributions can also be simulated even under the random bit model, in which one is only allowed to access an unbiased IID source.

On the other hand in the derivation of any \emph{nonlocal theorem} one crucial assumption is measurement independence: that measurement settings can be chosen independent of any underlying variables describing the system. Though in \cite{Shimony} Shimony \emph{et al.} have emphasized the reasonableness of this postulate, in recent time Hall \cite{Hall1,Hall2} and Barrett and Gisin \cite{Barrett2} independently studied this assumption more explicitly. In \cite{Brans}, Brans gave an explicit local and deterministic model for correlations between any two spin-1/2 particles where an underlying random variable fully determines not only the joint measurement outcomes, but also the associated measurement settings, i.e. there is no measurement independence at all. Interestingly, introducing a suitable measure of the degree of measurement independence, a possible candidate to quantify the amount of nonlocality, Hall showed that one does not need to relax measurement independence fully to obtain a no-signaling and deterministic model of the singlet state rather only 14\% relaxation suffices \cite{Hall1}. 

In this paper we consider 3-qubit GHZ quantum correlations and present no-signaling and deterministic protocol to simulate such nonlocal correlations by relaxing degree of measurement independence, partially. This problem is the straightforward next step after the 2-qubit singlet state. We show that $71.5\%$ relaxation of the degree of measurement independence is sufficient to obtain a no-signaling and deterministic model for all equatorial Von Neumann measurements on the 3-qubit GHZ state. Note that relaxation of measurement independence is higher for the GHZ state in comparison to the singlet state which is not counterintuitive as refutation of EPR argument by considering GHZ state is strikingly more direct than singlet state. Interestingly, we also find that to reproduce the expectation value for all equatorial Von Neumann measurements on the 3-qubit GHZ state one does not need to relax $71.5\%$ measurement independence, rather only $62.5\%$ relaxation is sufficient.

The paper is organized in the following way. In Section (\ref{ghz}) we briefly discuss the GHZ correlation and also the simulation protocol of equatorial GHZ correlation as introduced by Branciard and Gisin. In Section (\ref{simulation}) we present our result i.e  measurement dependent but no-signaling and deterministic simulation protocols for all equatorial Von Neumann measurements on the tripartite GHZ state. Section (\ref{conclusion}) contains discussions and conclusions.   

\section{Tripartite GHZ correlation}\label{ghz}
GHZ state was first introduced in \cite{GHZ}. The authors of \cite{GHZ} and then Mermin in \cite{Mermin} showed that the refutation of EPR argument by GHZ state is strikingly more direct than the one Bell's theorem provides for Bohm's version of EPR. The refutation is not only stronger — it is no longer statistical and can be accomplished in a single run. Besides this foundational importance GHZ state exhibits various applications in information theoretic processes, for example quantum secret sharing \cite{Mark}, entanglement broadcasting \cite{Tong}, simultaneous quantum secure direct communication \cite{Jin} etc. 

The form of three-qubit GHZ state held among three parties, say Alice, Bob and Charlie, looks:
\begin{equation}
|\psi\rangle_{GHZ}=\frac{1}{\sqrt{2}}(|000\rangle+|111\rangle).
\end{equation}
Here $|0\rangle$ ($|1\rangle$) represents the eigenvector of the Pauli Z operator with eigenvalue $+1$ ($-1$). Let the three parties perform spin measurements along the direction $\hat{m}_A$, $\hat{m}_B$ and $\hat{m}_C$, respectively; where the measurement direction for the party $X$ $(=A,B,C)$ is specified by the block vector $\hat{m}_X\equiv (\sin\theta_X\cos\phi_X,\sin\theta_X$ $\sin\phi_X, \cos\theta_X)$, with $\theta_X\in[0,\pi]$ and $\phi_X\in[0,2\pi]$. Denoting measurement outcome as $a$, $b$ and $c$ respectively, where $a,b,c\in[+1,-1]$, the three party GHZ correlation can be expressed in the following form:
\begin{eqnarray}\label{ghzcorre}
P(abc|\{\hat{m}_X\})&=&\frac{1}{8}[1+ab\cos\theta_A\cos\theta_B\nonumber\\
&&+bc\cos\theta_B\cos\theta_C+ca\cos\theta_C\cos\theta_A\nonumber\\
&&+abc\cos\theta_A\cos\theta_B\cos\theta_C\nonumber\\
&&+abc\sin\theta_A\sin\theta_B\sin\theta_C\nonumber\\
&&\cos(\phi_A+\phi_B+\phi_C)],
\end{eqnarray}
where $P(abc|\{\hat{m}_X\})\equiv P(abc|\hat{m}_A,\hat{m}_B,\hat{m}_C)$ is the probability of obtaining outcome $a$, $b$ and $c$ by Alice, Bob and Charlie, respectively, when they perform measurements $\hat{m}_A$, $\hat{m}_B$ and $\hat{m}_C$ on their respective parts of the shared GHZ state. If the measurement directions for all the parties are chosen from an equatorial plane (i.e $\sin\theta_A=\sin\theta_B=\sin\theta_C=0$) then Eqn.(\ref{ghzcorre}) becomes:
\begin{equation}\label{eq-ghzcorre}
P(abc|\{\hat{m}_X\})=\frac{1}{8}[1+abc\cos(\phi_A+\phi_B+\phi_C)].
\end{equation}
In this case the expectation value becomes:
\begin{eqnarray}\label{expt}
\langle\hat{m}_A\hat{m}_B\hat{m}_C\rangle&=&\sum_{a,b,c}abcP(abc|\hat{m}_A,\hat{m}_B,\hat{m}_C)\nonumber\\
&=&\cos(\phi_A+\phi_B+\phi_C)
\end{eqnarray}
while all single and bipartite marginals vanish. Note that although the choice of equatorial measurements is restrictive, these are enough to come up with the ``GHZ paradox''.
\subsection*{Branciard- Gisin simulation protocol}
\begin{figure}[h!]
   \includegraphics[width=8cm, height=4cm]{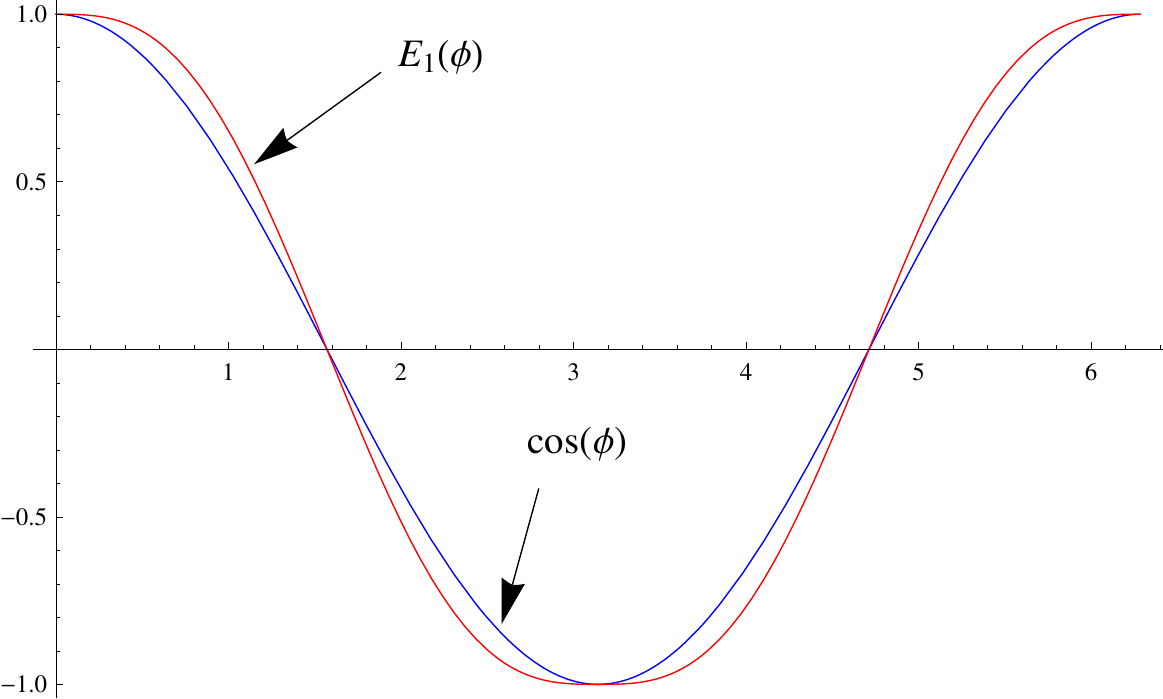}
    \caption{(Color on-line). Correlation $E_1(\phi)$ is the BG correlation whereas $\cos(\phi)$ is the desired GHZ correlation. This figure is taken from \cite{Branciard}.}
    \label{fig1}
\end{figure}

In \cite{Branciard}, Branciard and Gisin provided a simulation protocol for the correlation (\ref{expt}) with  bounded communications among the parties. They first showed that with 3 bits of classical communication (2 bits from Bob to Alice and 1 bit from Charlie to Alice) supplemented by several shared randomness and also local variables, Alice, Bob and Charlie can simulate the following statistics:
\begin{equation}\label{bg-corre}
E_1(\phi)=1-\frac{2\phi-\sin2\phi}{\pi}~~\mbox{for}~~\phi\in[0,\pi]
\end{equation}  
Of course Branciard-Gisin(BG) simulation protocol provides vanishing single and bipartite marginals. Moreover correlation $E_1(\phi)$ is stronger than the desired $\cos\phi$ correlation in the sense that $|E_1(\phi)|\ge|\cos\phi|$, for all $\phi$ (see Fig.\ref{fig1}). 
They then show that mixing the correlations of the form $E((2m+1)\phi)$, with $m\in Z$, one can obtain the desired $\cos\phi$ correlation as $E((2m+1)\phi)$ will preserve the perfect (anti-)correlations for $\phi=0$ and $\pi$. The mixing is done in the following way
\begin{equation}
\cos\phi=\sum_{m\ge 0}p_{2m+1}E((2m+1)\phi) 
\end{equation}
with $p_{2m+1}\ge 0$ for all $m\ge 0$. In particular, for $\phi=0$ one gets $\sum_{m\ge 0}p_{2m+1}=1$.
\section{Measurement dependence simulation protocol}\label{simulation}
In this section we provide a simulation protocol for the equatorial GHZ correlation with reduced measurement independence. Before providing the simulation protocol we briefly describe measurement (in)dependence.    

In simulation of nonlocal correlation, the parties involved are allowed to hold pre-shared variables, say $\lambda\in \Lambda$, $\Lambda$ being the space of shared variable(s) or the hidden variable(s). The assumption of measurement independence demands that distribution of the underlying variable is independent of the measurement settings, i.e.,
\begin{equation}\label{mi}
p(\lambda|\{\hat{m}_X\})=p(\lambda|\{\hat{m}'_X\}).
\end{equation}
Where $\hat{m}_X$ and $\hat{m}'_X$ are two different measurement settings for the party $X$. In the BG simulation Alice and Bob hold shared variables but distribution of these variables do not depend on the measurement settings $-$ satisfying the measurement independence assumption. Measurement independence is often justified by the notion of experimental free will, i.e., that experimenters can freely choose between different measurement settings irrespective of the underlying variable $\lambda$ describing the system. 

The degree to which an underlying model violates measurement independence is most simply quantified by the variational distance \cite{Hall1}:
\begin{equation}\label{measu}
M:=\sup_{\{\hat{m}_X\},\{\hat{m}'_X\}}\int d\lambda|p(\lambda|\{\hat{m}_X\})-
p(\lambda|\{\hat{m}'_X\})|
\end{equation}
with $0\le M\le2$. Clearly, a distance of $M=0$ corresponds to the case of full measurement independence as per Eqn.(\ref{mi}), consistent with maximum experimental free will in choosing measurement settings. Conversely, suppose that $M$ attains its greatest possible value, $M=2$, for some model. Hence, there are at least two particular joint measurement settings, $\{\hat{m}_X\}$ and $\{\hat{m}'_X\}$, such that for any $\lambda$ at most one of these settings is possible. Hence, no experimental free will whatsoever can be exercised to choose between these settings.

The fraction of measurement independence corresponding to a given model is defined by \cite{Hall1,Hall2}:
\begin{equation}
F:=1-\frac{M}{2}
\end{equation}
with $0\le F\le1$, where $F = 0$ corresponds to the case where no experimental free will can be enjoyed to choose measurement settings and $F = 1$ corresponds to complete experimental free will. Any value for $F$ strictly lying in between $0$ and $1$ corresponds to the case where experimenter's free will in choosing the measurement settings is restricted. Note that, geometrically, $F$ also represents the minimum degree of overlap between any two
underlying distributions $p(\lambda|\{\hat{m}_X\}$ and $p(\lambda|\{\hat{m}'_X\}$.
\subsection{Simulation protocol}  
Let Alice, Bob and Charlie share a variable $\vec{\lambda}$ chosen from unit circle. Now given a measurement direction from equatorial plane Alice, Bob and Charlie give their answers in the following way:
\begin{eqnarray}\label{assign}
A(\hat{m}_A,\vec{\lambda})=\mbox{sign}(\hat{m}_A.\vec{\lambda}),\nonumber\\
B(\hat{m}_B,\vec{\lambda})=\mbox{sign}(\hat{m}_B.\vec{\lambda}),\nonumber\\
C(\hat{m}_C,\vec{\lambda})=\mbox{sign}(\hat{m}_C.\vec{\lambda}),
\end{eqnarray}
where
\begin{eqnarray*}
\mbox{sign}(\alpha)=s(\alpha):=\left\{\begin{array}{ll}~~1~~\mbox{if $\alpha\geq 0$}\\~-1~~\mbox{otherwise}
\end{array}\right.
\end{eqnarray*}
The distribution of the variable $\vec{\lambda}$ is not uniform, rather the distribution depends on measurement directions of Alice, Bob and Charlie. The distribution is given by:    
\begin{eqnarray}\label{distri}
\rho(\vec{\lambda}|\{\hat{m}_X\}):=\rho'(\vec{\lambda}|\{\hat{m}_X\})\Theta(\phi_{AB}-\phi_{AC})\nonumber\\
+\rho''(\vec{\lambda}|\{\hat{m}_X\})\Theta(\phi_{AC}-\phi_{AB})
\end{eqnarray}
where $\Theta$ is the step function defined as:
\begin{eqnarray}
\Theta(\alpha)=\left\{\begin{array}{ll}~~1~~\mbox{if $\alpha\geq 0$}\\~~0~~\mbox{otherwise}
\end{array}\right.
\end{eqnarray}
$\phi_{AB}$ : angle between measurement directions of Alice and Bob;\\ 
$\phi_{AC}$ : angle between measurement directions of Alice and Charlie;\\
\begin{eqnarray}\label{distri1}
\rho'(\vec{\lambda}|\{\hat{m}_X\}):&=&\frac{1+\beta\cos (\phi_A+\phi_B+\phi_C)}{8(\pi-\phi_{AB})}\nonumber\\
&&\mbox{if $s(\hat{m}_A.\vec{\lambda})=s(\hat{m}_B.\vec{\lambda})=s(\hat{m}_C.\vec{\lambda})=\beta$}\nonumber\\
:&=&\frac{1+\beta\cos (\phi_A+\phi_B+\phi_C)}{8(\phi_{AB}-\phi_{AC})}\nonumber\\
&&\mbox{if $-s(\hat{m}_A.\vec{\lambda})=-s(\hat{m}_B.\vec{\lambda})=s(\hat{m}_C.\vec{\lambda})=\beta$}\nonumber\\
:&=&\frac{1+\beta\cos (\phi_A+\phi_B+\phi_C)}{8\phi_{AB}}\nonumber\\
&&\mbox{if $s(\hat{m}_A.\vec{\lambda})=-s(\hat{m}_B.\vec{\lambda})=-s(\hat{m}_C.\vec{\lambda})=\beta$}\nonumber\\
:&=&\frac{1+\beta\cos (\phi_A+\phi_B+\phi_C)}{8}\delta(\vec{\lambda}-\beta\vec{\lambda}_0)\nonumber\\
&&\mbox{if $-s(\hat{m}_A.\vec{\lambda})=s(\hat{m}_B.\vec{\lambda})=-s(\hat{m}_C.\vec{\lambda})=\beta$}\nonumber\\
\end{eqnarray}
with $\beta\in\{+1,-1\}$; and 
\begin{eqnarray}
\rho''(\vec{\lambda}|\{\hat{m}_X\}):&=&\frac{1+\beta\cos (\phi_A+\phi_B+\phi_C)}{8(\pi-\phi_{AC})}\nonumber\\
&&\mbox{if $s(\hat{m}_A.\vec{\lambda})=s(\hat{m}_B.\vec{\lambda})=s(\hat{m}_C.\vec{\lambda})=\beta$}\nonumber\\
:&=&\frac{1+\beta\cos (\phi_A+\phi_B+\phi_C)}{8}\delta(\vec{\lambda}-\beta\vec{\lambda}_0)\nonumber\\
&&\mbox{if $-s(\hat{m}_A.\vec{\lambda})=-s(\hat{m}_B.\vec{\lambda})=s(\hat{m}_C.\vec{\lambda})=\beta$}\nonumber\\
:&=&\frac{1+\beta\cos (\phi_A+\phi_B+\phi_C)}{8\phi_{AC}}\nonumber\\
&&\mbox{if $s(\hat{m}_A.\vec{\lambda})=-s(\hat{m}_B.\vec{\lambda})=-s(\hat{m}_C.\vec{\lambda})=\beta$}\nonumber\\
:&=&\frac{1+\beta\cos (\phi_A+\phi_B+\phi_C)}{8(\phi_{AC}-\phi_{AB})}\nonumber\\
&&\mbox{if $-s(\hat{m}_A.\vec{\lambda})=s(\hat{m}_B.\vec{\lambda})=-s(\hat{m}_C.\vec{\lambda})=\beta$}\nonumber\\
\end{eqnarray}
$\vec{\lambda_0}$ be a fixed vector chosen from the unit circle and $\delta$ be the Dirac delta function. Given the value assignment as Eqn.(\ref{assign}) and the distribution of the hidden variable $\vec{\lambda}$ as Eqn.(\ref{distri}), it is easy to check that the quantum probability distribution of Eqn.(\ref{eq-ghzcorre}) is obtained by integrating out the variable $\vec{\lambda}$, i.e.,
\begin{equation}
P(abc|\{\hat{m}_X\})=\int d\vec{\lambda} p(abc|\vec{\lambda})\rho(\vec{\lambda}|\{\hat{m}_X\}),
\end{equation}
where $p(abc|\vec{\lambda})=\delta_{a,A(\hat{m}_A,\vec{\lambda})}\delta_{b,B(\hat{m}_B,\vec{\lambda})}\delta_{c,C(\hat{m}_C,\vec{\lambda})}$. The amount of degree of measurement dependence for the above simulation protocol turns out to be $M\simeq 1.43$ and hence $F\simeq 28.5\%$ (Appendix \ref{calcu}). Thus to obtain a deterministic, no-signaling model for the equatorial Von Neumann measurements on tripartite GHZ state one does not need to give up complete experimental free-will, rather $71.5\%$ lack of measurement independence is sufficient. Note that, if in some simulation protocol of a correlation the fraction of measurement independence takes a value $0<f<1$, it does not imply that there exist a simulation model that replicates the corresponding correlation as long as one can have a dependent model at least $1-f$ percent of the time but must have an independent model the other fraction $f$ of time. This is because the definition of $F$ lacks such an operational interpretation. 

It is noteworthy that given an expectation value the probability distribution associated with it is not unique. Interestingly we find that if one wants to reproduce the expectation value of the equatorial measurements on GHZ state i.e. Eqn.(\ref{expt}) but does not bother about reproducing the quantum probabilities (i.e Eqn.(\ref{eq-ghzcorre})) then there exists a simulation model where the amount of experimental free-will that has to be sacrificed is $62.5\%$(Appendix \ref{expt-simu}). 
\section{Discussion and conclusion}\label{conclusion}  
In this paper we provide a protocol based upon relaxation of measurement independence which can reproduce the equatorial Von Neumann measurements results on three qubits GHZ deterministically and unlike ref.\cite{Branciard} our simulation protocol uses no classical communications among the parties. This result tells that if sufficient amount of \emph{free-will} is not assured for the involved parties to chose their measurement settings then GHZ argument cannot reveal the conflict between quantum theory and local realism. In \cite{Hall2}, the author has shown that EPR-Kochen-Specker theorem due to Mermin \cite{Mermin} fails if measurement independence is relaxed by $33.3\%$. Please note that, neither any local deterministic simulation protocol for GHZ state has been given in \cite{Hall2} nor it is known whether $33.3\%$ relaxation of measurement independence is sufficient for modeling equatorial Von Neumann measurements statistics locally and deterministically. Thus optimality of our model is a question of further investigation. It is also interesting to extend this model to all measurements (which in our case is restricted to equatorial Von Neumann measurements). 

In the bipartite scenario it has been shown in \cite{Manik1} that it is possible to construct a local deterministic simulation protocol for singlet correlation by sacrificing measurement independence of one party whereas the other party enjoys complete free-will. In reduced measurement independence simulation of GHZ correlation it is also interesting to study whether such a model is possible where some (and not all ) parties sacrifice measurement independence and rest enjoy complete free-will.

\appendix\normalsize
\section{Appendix A: Calculation of measurement dependency}\label{calcu}
\begin{figure}[t!]
\begin{center}
\includegraphics[width=7cm, height=5cm]{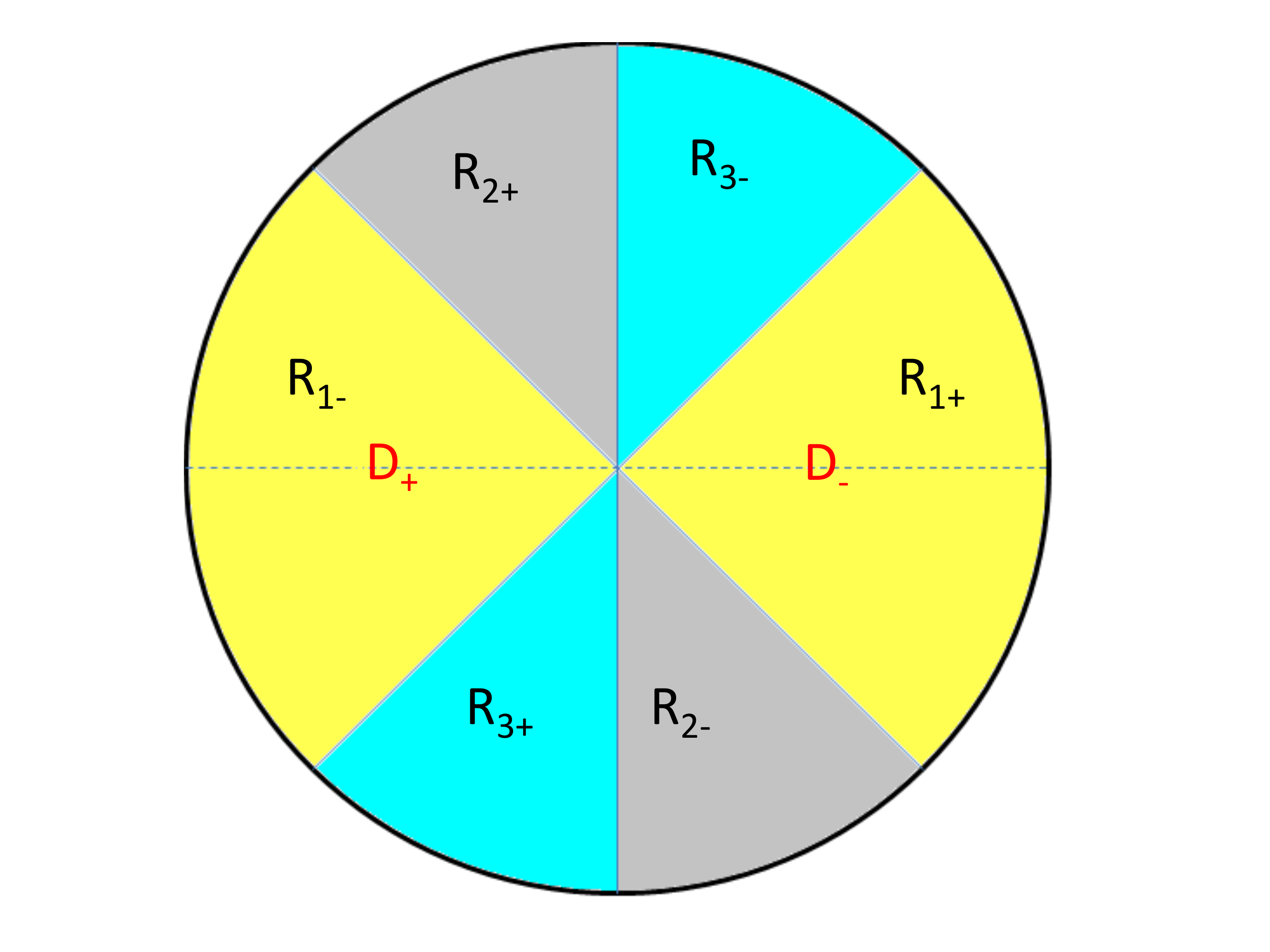}
\end{center}
\caption{(Color on-line). $R_{1+} (R_{1-})$ is the region where $s(\hat{m}_A.\vec{\lambda})=s(\hat{m}_B.\vec{\lambda})=s(\hat{m}_C.\vec{\lambda})=+1 (-1)$; $R_{2+} (R_{2-})$ is the region where $s(\hat{m}_A.\vec{\lambda})=-s(\hat{m}_B.\vec{\lambda})=-s(\hat{m}_C.\vec{\lambda})=+1 (-1)$ and $R_{3+} (R_{3-})$ is the region where $-s(\hat{m}_A.\vec{\lambda})=-s(\hat{m}_B.\vec{\lambda})=s(\hat{m}_C.\vec{\lambda})=+1 (-1)$. $D_+$ ($D_{-}$) denotes the fixed vector $\vec{\lambda_0}$ (-$\vec{\lambda_0}$)} 
\label{pic}
\end{figure}

\begin{figure}[t!]
 \begin{center}
   \includegraphics[width=5cm, height=5cm]{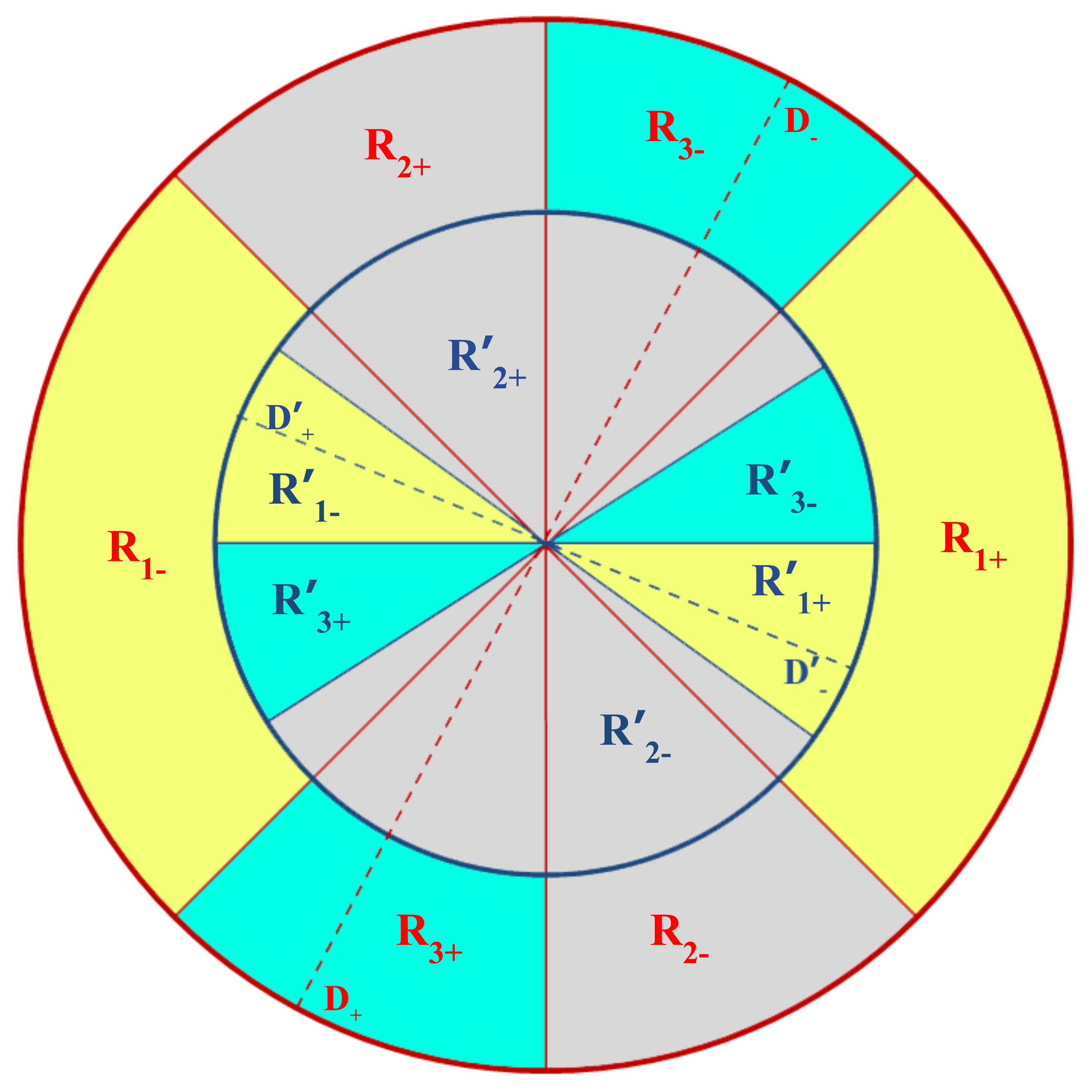}
   \end{center}
    \caption{(Color on-line). This figure shows distribution of HV corresponding to measurement settings $\{\hat{m}_X\}$ and $\{\hat{m}'_X\}$. The inner circle represents the distribution corresponding to the measurement setting $\{\hat{m}'_X\}$ and the outer circle corresponding to the measurement setting $\{\hat{m}_X\}$. The maximum of the right hand side of Eqn.(\ref{measu}) occurs when $R_{1+}$ contains  $D'_-$ and the region $R'_{3-}$ completely and the region $R'_{2-}$ partially.}
    \label{pic10}
\end{figure}

The degree of measurement dependency is quantified as Eqn.(\ref{measu}). To find $M$ we have to maximize the difference between the densities of hidden variable (HV) for all pairs of measurement settings over the HV space. The difference will be maximized if the upper value of the distribution of HV corresponding to one measurement setting overlaps by maximum amount with the lower value of the other measurement setting. The distribution of HV for each measurement setting consists of three pair of regions comprising two opposite spherical sectors (\{$R_{1+},R_{1-}$\},\{$R_{2+},R_{2-}$\},\{$R_{3+},R_{3-}$\}) and a pair of delta functions ($D_+,D_-$) as defined in Eqn.(\ref{distri1})and shown in Fig.\ref{pic}. 
The distribution of HV for measurement setting $\{\hat{m}_X\}$ and $\{\hat{m}'_X\}$ are denoted by prime and unprimed region, respectively (see Fig.\ref{pic10}). The maximum of right hand side of Eqn.(\ref{measu}) occurs when  $R_{1+}$ contains  $D'_-$ and the region $R'_{3-}$ completely and the region $R'_{2-}$ partially as shown in Fig.\ref{pic10}.
It becomes that $M=1.43$ and thus we have $F:=1-\frac{M}{2}\simeq 28.5\%$.
\section{Appendix B: Simulation of GHZ expectation value}\label{expt-simu}
In this case also Alice, Bob and Charlie share a variable $\vec{\lambda}$ chosen from unit circle and given a measurement direction from equatorial plane Alice, Bob and Charlie give there answer like Eqn.(\ref{assign}). The distribution of the variable $\vec{\lambda}$ in this case is given by
\begin{eqnarray*}\label{distri-stat}
\vartheta(\vec{\lambda}|\{\hat{m}_X\}):=\vartheta'(\vec{\lambda}|\{\hat{m}_X\})\Theta(\phi_{AB}-\phi_{AC})\\
+\vartheta''(\vec{\lambda}|\{\hat{m}_X\})\Theta(\phi_{AC}-\phi_{AB})
\end{eqnarray*}
where 
\begin{eqnarray*}
\vartheta'(\vec{\lambda}|\{\hat{m}_X\}):&=&\frac{1+\beta\cos (\phi_A+\phi_B+\phi_C)}{6(\pi-\phi_{AB})}\nonumber\\
&&\mbox{if $s(\hat{m}_A.\vec{\lambda})=s(\hat{m}_B.\vec{\lambda})=s(\hat{m}_C.\vec{\lambda})=\beta$}\nonumber\\
:&=&\frac{1+\beta\cos (\phi_A+\phi_B+\phi_C)}{6(\phi_{AB}-\phi_{AC})}\nonumber\\
&&\mbox{if $-s(\hat{m}_A.\vec{\lambda})=-s(\hat{m}_B.\vec{\lambda})=s(\hat{m}_C.\vec{\lambda})=\beta$}\nonumber\\
:&=&\frac{1+\beta\cos (\phi_A+\phi_B+\phi_C)}{6\phi_{AB}}\nonumber\\
&&\mbox{if $s(\hat{m}_A.\vec{\lambda})=-s(\hat{m}_B.\vec{\lambda})=-s(\hat{m}_C.\vec{\lambda})=\beta$}\nonumber\\
\end{eqnarray*}
with $\beta\in\{+1,-1\}$; and 
\begin{eqnarray*}
\vartheta''(\vec{\lambda}|\{\hat{m}_X\}):&=&\frac{1+\beta\cos (\phi_A+\phi_B+\phi_C)}{6(\pi-\phi_{AC})}\nonumber\\
&&\mbox{if $s(\hat{m}_A.\vec{\lambda})=s(\hat{m}_B.\vec{\lambda})=s(\hat{m}_C.\vec{\lambda})=\beta$}\nonumber\\
:&=&\frac{1+\beta\cos (\phi_A+\phi_B+\phi_C)}{6\phi_{AC}}\nonumber\\
&&\mbox{if $s(\hat{m}_A.\vec{\lambda})=-s(\hat{m}_B.\vec{\lambda})=-s(\hat{m}_C.\vec{\lambda})=\beta$}\nonumber\\
:&=&\frac{1+\beta\cos (\phi_A+\phi_B+\phi_C)}{6(\phi_{AC}-\phi_{AB})}\nonumber\\
&&\mbox{if $-s(\hat{m}_A.\vec{\lambda})=s(\hat{m}_B.\vec{\lambda})=-s(\hat{m}_C.\vec{\lambda})=\beta$}\nonumber\\
\end{eqnarray*}
In this case it becomes $F=37.5\%$. Therefore $62.5\%$ lack of measurement independence for each party is sufficient to simulate statistic of the equatorial Von Neumann measurements on GHZ state.

\textbf{Acknowledgment}: It is a pleasure to thank Guruprasad Kar for many stimulating discussions. MB acknowledge C. Branciard for fruitful discussions. AM acknowledges support from CSIR, Govt. of India (File No.09/093(0148)/ 2012-EMR-I).

\end{document}